\documentclass[letter,twocolumn]{jpsj3}
\usepackage{txfonts}
\usepackage{bm}
\usepackage{braket}
\usepackage{color}

\def\MAG{\mathrm{mag}}
\def\E{\mathrm{e}}
\def\QQ{{\bm q}}
\def\D{\mathrm{d}}
\def\KB{k_\mathrm{B}}
\def\KK{{\bm k}}
\def\I{\mathrm{i}}
\def\rr{\mathrm{r}}
\def\aa{\mathrm{a}}
\def\KKF{{\bm k_{\mathrm{F}\sigma}}}

\title{Microscopic Theory of Magnon-Drag Thermoelectric Transport in Ferromagnetic Metals}

\author{\name{Daisuke \surname{Miura}}\thanks{E-mail address: dmiura@solid.apph.tohoku.ac.jp} and \name{Akimasa \surname{Sakuma}}}
\inst{
\address{Department of Applied Physics, Tohoku University,\\Sendai 980-8579}
}
\abst{
A theoretical study of the magnon-drag Peltier and Seebeck effects in ferromagnetic metals is presented.
A magnon heat current is described perturbatively from the microscopic viewpoint with respect to electron--magnon interactions and the electric field.
Then, the magnon-drag Peltier coefficient $\Pi_\MAG$ is obtained as the ratio between the magnon heat current and the electric charge current.
We show that $\Pi_\MAG=C_\MAG T^{5/2}$ at a low temperature $T$; that the coefficient $C_\MAG$ is proportional to the spin polarization $P$ of the electric conductivity;
and that $P>0$ for $C_\MAG<0$, but $P<0$ for $C_\MAG>0$.
From experimental results for magnon-drag Peltier effects, we estimate that the strength of the electron--magnon interaction is about $0.3$ eV$\cdot\AA^{3/2}$ for permalloy.
}

\kword{Peltier effect, Seebeck effect, magnon drag, spin caloritronics, spintronics, spin current, electron--magnon interaction}

\begin{document}
\maketitle
Understanding interactions between spin dynamics and transport phenomena in ferromagnetic materials is one of the central issues in spintronics and spin caloritronics\cite{spincaloritronics}. In particular, it is important
to clarify the thermal properties of spin dynamics, that is, magnon scattering mechanisms, at finite temperature.
The magnon-drag thermoelectric effects proposed by Bailyn\cite{PhysRev.126.2040} are good subjects for
investigating the thermal properties and transport phenomena of magnons
because the effects are determined by substance-specific properties\cite{MacDonald} and are sensitive to external magnetic fields.
Moreover, the effects are useful in the study of fundamental electron--magnon interactions.
Blatt {\it et al.}\cite{PhysRevLett.18.395} reported that the Seebeck coefficient $S$ for iron takes a maximum value near 200 K, that $S$ is unresponsive to doping with heavy atoms and annealing, and that $S$ is well fitted by the form $S=C_1 T+ C_2 T^{3/2}$ (where $C_1$ and $C_2$ are constants).
These differ from the features expected for phonon-drag Seebeck effects\cite{MacDonald} both quantitatively and qualitatively.
Grannemann and Berger\cite{PhysRevB.13.2072} phenomenologically derived a convenient expression for the magnon-drag Peltier coefficient in ferromagnetic metals,
assuming that the drift velocity of magnons, $\bm v_\MAG$, is proportional to the drift velocity of electrons (i.e., $\bm v_\MAG=\eta\bm v_\E$),
\begin{align}
\Pi_\MAG^\eta=\frac{2\eta D}{3en_\E V}\sum_\QQ \QQ^2\omega_\QQ\frac{\D n(\omega_\QQ)}{\D\omega_\QQ},\label{eq1}
\end{align}
where $D$ is the spin-wave stiffness constant, $e>0$ is the elementary charge, $n_\E$ is the number of conduction electrons per unit volume,
$V$ is the volume of the system,
$n(x):=[\exp(x/\KB T)-1]^{-1}$ is the Bose--Einstein distribution function, $\KB$ is the Boltzmann constant, $T$ is the absolute temperature, and $\omega_\QQ$ is the energy dispersion relation for magnons.
By using this expression, they found that at low temperature, $\Pi_\MAG^\eta\propto T^{5/2}$, and that consequently, the magnon-drag Seebeck coefficient $S_\MAG^\eta=\Pi_\MAG^\eta/T\propto T^{3/2}$.
In addition, by fitting eq. (\ref{eq1}) to experimental data, they obtained the values $\eta=2.98$ for Ni$_{69}$Fe$_{31}$ and $\eta=2.20$ for Ni$_{66}$Cu$_{34}$.
Recently, Costache {\it et al.}\cite{Costache} have developed a magnon-drag thermopile
that can cancel out all contributions except for that of magnons to thermopower,
allowing the electric voltage induced by the magnon-drag Seebeck effect to be measured directly.
They found that the Seebeck coefficient of a magnon-drag thermopile made of permalloy takes a peak near 180 K and is well fitted to $\Pi_\MAG^\eta/ T$ at temperatures below 180 K ($\eta=3$).
As above, the phenomenological expression (\ref{eq1}) can accurately describe the experimental results. However, there has been no progress on the theoretical front for understanding magnon-drag thermoelectric effects from the microscopic viewpoint.

In this letter, we describe magnon-drag thermoelectric phenomena in ferromagnetic metals at the microscopic level and aim to obtain detailed information on the magnon-drag thermoelectric coefficients.
We calculate a magnon thermal current driven by an electric field using the Keldysh Green function technique.
The mathematical form of the resultant magnon-drag Peltier coefficient conforms to Grannemann and Berger's expression (\ref{eq1}) and a microscopic expression for $\eta$ is obtained. Furthermore, we show that $\eta$, a physically important feature, is proportional to the spin polarization of the electric conductivity.

\begin{figure}[htbp]
\includegraphics[width=0.5\textwidth]{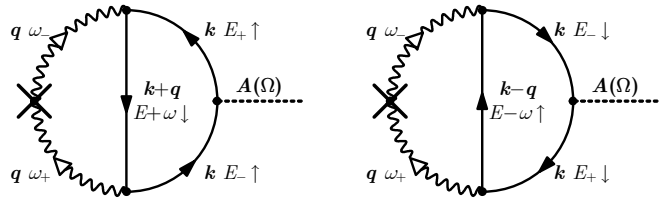}
\caption{Diagrammatic representations of magnon-drag Peltier effect. The solid and wavy lines represent the Keldysh Green functions of the electron and magnon, respectively; the dashed line stands for the vector potential; and the cross stands for
the magnon heat current vertex.
}
\label{fig1}
\end{figure}

Let us consider a Hamiltonian describing ferromagnetic metals $\mathcal{H}:=\sum_{\KK\sigma}\varepsilon_{\KK\sigma}c_{\KK\sigma}^\dagger c_{\KK\sigma}^{}+\sum_{\QQ}\omega_\QQ b^\dagger_{\QQ} b^{}_\QQ+\mathcal{H}_\mathrm{em}$,
where $c_{\KK\sigma}^{}$ is an operator that annihilates the $\sigma$ spin electron with an wave vector $\KK$, $\varepsilon_{\KK\sigma}:=\hbar^2\KK^2/2m-\sigma\Delta/2-\mu$ is the energy dispersion relation for electrons calculated from the chemical potential $\mu$, $m$ is the mass of an electron, $\Delta$ stands for the exchange splitting, $b_\QQ$ is an operator that annihilates a magnon with an wave vector $\QQ$, and $\mathcal{H}_\mathrm{em}$ is an electron--magnon interaction defined by\cite{PTP.16.58,PhysRev.163.503,PhysRevB.17.3666}
\begin{align*}
\mathcal{H}_\mathrm{em}:=\frac{I}{\sqrt{V}}\sum_{\KK,\QQ}\Biggl[
b_\QQ^\dagger c_{\KK\uparrow}^\dagger c_{\KK+\QQ\downarrow}^{}
+
b_\QQ c_{\KK+\QQ\downarrow}^\dagger c_{\KK\uparrow}^{}
\Biggr],
\end{align*}
where $I$ represents the strength of the electron--magnon interaction.
The interaction with a static homogeneous electric field $\bm E:=-\D\bm A(t)/\D t$ is given by $\mathcal{H}'(t):=-\bm j_\E\cdot\bm A(t)V$ in terms of the electric charge current density operator $\bm j_\E:=-(e\hbar/mV)\sum_{\KK\sigma}\KK c_{\KK\sigma}^\dagger c_{\KK\sigma}^{}$.
We define the space-averaged magnon heat current density operator by
\begin{align*}
\bm j_\MAG(t):=\frac{2D}{\hbar V}\sum_\QQ \QQ\omega_\QQ b_\QQ^\dagger(t) b_\QQ^{}(t),
\end{align*}
where operators are in the Heisenberg representation with respect to $\mathcal{H}+\mathcal{H}'(t)$.
Using the lesser function\cite{Haug,Rammer} of magnons
$D^<_\QQ(t,t'):=-(\I/\hbar)\Braket{b_\QQ^\dagger(t') b_\QQ^{}(t)}$,
the statistical average of $\bm j_\MAG(t)$ can be represented by
\begin{align*}
\Braket{\bm j_\MAG(t)}=\frac{\I 2D}{V}\sum_\QQ \QQ\omega_\QQ D^<_\QQ(t,t),
\end{align*}
where $\Braket{\cdots}$ is a statistical average in $\mathcal{H}+\mathcal{H}'(t)$ when $\mathcal{H}'(t)$ is regarded as the nonequilibrium part.
We perform the lowest-order calculation for $\Braket{\bm j_\MAG(t)}$ with respect to both $\mathcal{H}_\mathrm{em}$ and $\mathcal{H}'(t)$ as shown in Fig. \ref{fig1}.
In energy space, we have
\begin{align*}
\Braket{\bm j_\MAG(t)}\simeq\frac{\I 2 D}{V}\sum_\QQ \QQ\omega_\QQ\int\frac{\D\Omega}{2\pi\hbar}\E^{-\I\Omega t/\hbar}\int\frac{\D\omega}{2\pi\hbar}\varDelta D_\QQ^<(\omega_+,\omega_-),
\end{align*}
where $\varDelta D_\QQ^<(\omega_+,\omega_-)$ corresponds to the diagrams shown in Fig. \ref{fig1} and is given by
\begin{align*}
\varDelta D^<_\QQ(\omega_+,\omega_-)&:=-\frac{\I e\hbar I^2}{2\pi m}\bm A(\Omega)\cdot d_\QQ(\omega_+)\bm \Lambda_\QQ(\omega_+,\omega_-)d_\QQ(\omega_-)\Biggr|^<,\\
\bm \Lambda_\QQ(\omega_+,\omega_-)&:=\frac{1}{V}\sum_\KK\KK\int\D E\Biggl[
g_{\KK\uparrow}(E_+)g_{\KK\uparrow}(E_-)g_{\KK+\QQ\downarrow}(E+\omega)
\notag\\
&
+
g_{\KK\downarrow}(E_+)g_{\KK\downarrow}(E_-)g_{\KK-\QQ\uparrow}(E-\omega)
\Biggr],
\end{align*}
where $\omega_\pm:=\omega\pm\Omega/2$, $E_\pm:=E\pm\Omega/2$, $\bm A(\Omega):=\int\D t\ \E^{\I\Omega t/\hbar}\bm A(t)$, and $g_{\KK\sigma}(E)$ and $d_\QQ(\omega)$ are
the unperturbed Keldysh Green functions of the $\sigma$ spin electron with the wave vector $\KK$ and the magnon with the wave vector $\QQ$, respectively.
Taking the lesser component, we obtain $d_\QQ(\omega_+)\bm \Lambda_\QQ(\omega_+,\omega_-)d_\QQ(\omega_-)|^<
=
d_\QQ^\rr(\omega_+)\bm \Lambda_\QQ^\rr(\omega_+,\omega_-)d_\QQ^<(\omega_-)
+
d_\QQ^\rr(\omega_+)\bm \Lambda_\QQ^<(\omega_+,\omega_-)d_\QQ^\aa(\omega_-)
+
d_\QQ^<(\omega_+)\bm \Lambda_\QQ^\aa(\omega_+,\omega_-)d_\QQ^\aa(\omega_-)
$
where $d_\QQ^<(\omega)=-n(\omega)[d_\QQ^\aa(\omega)-d_\QQ^\rr(\omega)]$, and
$d_\QQ^\aa(\omega)$ and $d_\QQ^\rr(\omega)$ are the unperturbed advanced and retarded Green's functions of the magnon, respectively.
The part representing the electron--hole pair is, in the first order of $\Omega$, calculated as
\begin{align*}
\bm\Lambda^\rr_\QQ(\omega_+,\omega_-)&=-\frac{2\pi^2\Omega}{V}\sum_\KK\KK\int\D E\frac{\D f(E)}{\D E}\Biggl[
\rho_{\KK\uparrow}(E)^2 g_{\KK+\QQ\downarrow}^\rr(E+\omega)
\notag\\
&
+
\rho_{\KK\downarrow}(E)^2 g_{\KK-\QQ\uparrow}^\aa(E-\omega)
\Biggr],
\\
\bm\Lambda^\aa_\QQ(\omega_+,\omega_-)&=\bm\Lambda^\rr_\QQ(\omega_+,\omega_-)^*,
\\
\bm\Lambda^<_\QQ(\omega_+,\omega_-)&=\I\Im\frac{4\pi^2\Omega}{V}\sum_\KK\KK\int\D E\frac{\D f(E)}{\D E}
\notag\\
&
\times\Biggl[
f(E+\omega)
\rho_{\KK\uparrow}(E)^2 g_{\KK+\QQ\downarrow}^\rr(E+\omega)
\notag\\
&
+
\{
1-f(E-\omega)
\}
\rho_{\KK\downarrow}(E)^2 g_{\KK-\QQ\uparrow}^\aa(E-\omega)
\Biggr],
\end{align*}
where $f(x):=[\exp(x/\KB T)+1]^{-1}$ is the Fermi--Dirac distribution function, $g_{\KK\sigma}^\rr(E):=[E-\varepsilon_{\KK\sigma}+\I\delta]^{-1}=g_{\KK\sigma}^\aa(E)^*$, and $\rho_{\KK\sigma}(E):=-\Im g_{\KK\sigma}^\rr(E)/\pi$.
Note that $\bm\Lambda_\QQ(\omega,\omega)$ vanishes identically when performing $\bm k$ summation and $E$ integral.
The factor $-\D f(E)/\D E$ acts on
the Green's functions of the electron
approximately like the Dirac's delta function $\delta(E)$ under the condition $\delta\gg \KB T$.
This approximation is valid at temperatures below about 100 K because $\delta$ is on the order of $10^{-2}$ eV in metals.
Considering only the contribution of the $\sigma$ spin electrons to the Fermi surface $|\KK|=|\KKF|$ and the long-wavelength region $\hbar^2\KKF\cdot\QQ/m(\omega-\Delta)\ll 1$,
we obtain
\begin{align}
\bm\Lambda^\rr_\QQ(\omega_+,\omega_-)&\simeq\QQ\Omega\frac{2\pi m}{e^2\hbar}\frac{\sigma_\uparrow-\sigma_\downarrow}{(\omega-\Delta+\I\delta)^2},\label{eq3}\\
\bm \Lambda^<_\QQ(\omega_+,\omega_-)&\simeq\I 2\Biggl[\frac{\D}{\D\omega}\omega n(\omega)\Biggr]\Im \bm\Lambda^\rr_\QQ(\omega_+,\omega_-),\label{eq4}
\end{align}
where $\sigma_\sigma:=(\pi e^2\hbar^3/3m^2V)\sum_\KK\KK^2\rho_{\KK\sigma}(0)^2$ is the spin-dependent electric conductivity\cite{Mahan}, and we use the identities $\int\D E\ \D f(E)/\D E=-1$ and $\int\D E\ f(E+\omega)\D f(E)/\D E=\int\D E[1-f(E-\omega)]\D f(E)/\D E=(\D/\D\omega)\omega n(\omega)$.
Using eqs. (\ref{eq3}) and (\ref{eq4}), we have
\begin{align*}
\Braket{\bm j_\MAG(t)}
&\simeq
\bm j_\mathrm{c}
\frac{2I^2DP}{3\pi eV}\sum_\QQ \QQ^2\omega_\QQ\int\D\omega
\Im
\frac{1}{(\omega-\Delta+\I\delta)^2}
\notag\\
&
\times
\Biggl[
n(\omega)d_\QQ^\rr(\omega)^2
+
\omega\frac{\D n(\omega)}{\D\omega}
d_\QQ^\rr(\omega)
d_\QQ^\aa(\omega)
\Biggr],
\end{align*}
where $\bm j_\mathrm{c}:=(\sigma_\uparrow+\sigma_\downarrow)\bm E$ is the electric charge current and $P:=(\sigma_\uparrow-\sigma_\downarrow)/(\sigma_\uparrow+\sigma_\downarrow)$ is the spin polarization of the electric conductivity.
As a result, we obtain the magnon-drag Peltier coefficient
$\Pi_\MAG:=\Braket{\bm j_\MAG(t)}^\gamma/j_\mathrm{c}^\gamma$
($\gamma=x,y,z$). This result indicates that the magnon-drag Peltier coefficient is proportional to the spin polarization.
To evaluate the $\omega$ integral, we express the self-energy of the magnon in terms of the Gilbert damping constant $\alpha$ as\cite{PhysRevB.79.094415}
\begin{align*}
d_\QQ^\rr(\omega)=\frac{1}{\omega-\omega_\QQ+\I\alpha\omega}=d_\QQ^\aa(\omega)^*.
\end{align*}
Considering a low-energy region $\omega_\QQ<\KB T$ in the $\QQ$ summation and assuming $\alpha\ll 1$ and $\Delta\ll\delta/\alpha$, we obtain
\begin{align}
\Pi_\MAG\simeq
\frac{4I^2DP\delta\Delta}{3\alpha e(\Delta^2+\delta^2)^2V}\sum_\QQ\QQ^2\omega_\QQ\frac{\D n(\omega_\QQ)}{\D\omega_\QQ}.\label{eq5}
\end{align}
A comparison between eqs. (\ref{eq5}) and (\ref{eq1}) affords the following microscopic expression for the phenomenological parameter $\eta$:
\begin{align}
\eta=\frac{2I^2n_\E\delta\Delta}{\alpha(\Delta^2+\delta^2)^2}P. \label{eq6}
\end{align}
Grannemann and Berger\cite{PhysRevB.13.2072} have experimentally obtained $\eta=2.98$ for Ni$_{69}$Fe$_{31}$ using $n_\E=3.1\times 10^{-2}\AA^{-3}$ in eq. (\ref{eq1}).
Putting these values into eq. (\ref{eq6}), we can estimate
the strength of the electron--magnon interaction to be $I=0.3$ eV$\cdot\AA^{3/2}$ by using typical values for ferromagnetic metals: $\Delta=0.1$ eV\cite{kota}, $P=0.5$\cite{kota}, $\delta=0.01$ eV\cite{JPSJ.81.084701}, and $\alpha=0.01$\cite{JPSJ.81.084701}.

Equation (\ref{eq5}) could also be useful for determining the sign of $P$ by a comparison with the experimental data.
When $\omega_\QQ=D\QQ^2+\Delta_\mathrm{gap}$,
eq. (\ref{eq5}) can be written in the form
\begin{align*}
\Pi_\MAG&=-\frac{I^2\delta\Delta\KB^{5/2}}{3\pi^2\alpha eD^{3/2}(\Delta^2+\delta^2)^2}PT^{5/2}
\notag\\
&
\times
\int_0^\infty\D x\ x^{3/2}\Biggl(
x+\frac{\Delta_\mathrm{gap}}{\KB T}
\Biggr)
\frac{\exp(x+\Delta_\mathrm{gap}/\KB T)}{[\exp(x+\Delta_\mathrm{gap}/\KB T)-1]^2}\notag\\
&=:C_\MAG T^{5/2}
.
\end{align*}
Since $C_\MAG$ can be determined experimentally by fitting a function with respect to $T$ to the temperature dependence of the Peltier coefficient,
we can evaluate the sign of $P$, that is, $P>0$ for $C_\MAG<0$ and $P<0$ for $C_\MAG>0$.
This is analogous to the situation where the type of charge carriers in semiconductors can be determined by Peltier or Seebeck measurements.

Finally, we refer to a microscopic description for Seebeck effects.
The magnon-drag Seebeck effect is the counter phenomenon of the magnon-drag Peltier effect.
Therefore, it is expected that the two effects will satisfy Onsager's reciprocity relation, that is, the magnon-drag Seebeck coefficient must be given by $S_\MAG=\Pi_\MAG/T$.
The first work to microscopically describe a response to a temperature gradient was that of Luttinger\cite{PhysRev.135.A1505}.
Following that approach, we can describe Seebeck effects by introducing a Hamiltonian $\int\D\bm r h(\bm r)\phi_\mathrm{L}(\bm r,t)$, where $h(\bm r)$ is a total Hamiltonian density operator and $\phi_\mathrm{L}(\bm r,t)$ is a pseudoscalar potential.
In other words, we calculate the response of the nonequilibrium electric charge current to the Hamiltonian, and then replace $\bm\nabla\phi_\mathrm{L}(\bm r,t)$ with $(\bm\nabla T)/T$.
However, here we consider the Hamiltonian $-\bm j_\MAG\cdot\bm A_\mathrm{L}(t)V$ instead of $\int\D\bm r h(\bm r)\phi_\mathrm{L}(\bm r,t)$ from an analogy with the fact that the product of the charge density and the scalar potential corresponds to the product of the charge current density and the vector potential. Thus, we regard
\begin{align*}
-\bm j_\MAG\cdot\bm A_\mathrm{L}(t)V=-\frac{2D}{\hbar}\sum_\QQ\QQ\omega_\QQ b_\QQ^\dagger b_\QQ\cdot\bm A_\mathrm{L}(t)
\end{align*}
as a Hamiltonian describing a static homogeneous temperature gradient,
and we assume that $\D \bm A_\mathrm{L}(t)/\D t$ corresponds to $(\bm\nabla T)/T$ as $\bm E$ is given by $-\D\bm A(t)/\D t$.
This Hamiltonian has some advantages:
First,
the diffusion ladder arising from short-range impurity scattering vanishes for a static homogeneous temperature gradient.
Second,
it is easy to see that the Seebeck coefficient is associated with a charge current--energy current correlation function\cite{PTP.46.757}
because $-\bm j_\MAG\cdot\bm A_\mathrm{L}(t)V$ has the same form as the Hamiltonian $-\bm j_\E\cdot\bm A(t)V$ that we used to account for the static homogeneous electric field.

\begin{figure}[htbp]
\includegraphics[width=0.5\textwidth]{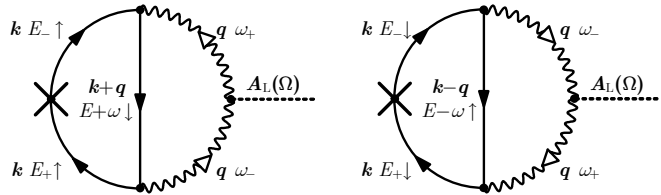}
\caption{Diagrammatic representations of magnon-drag Seebeck effect. The solid and wavy lines represent the Keldysh Green functions of the electron and magnon, respectively; the dashed line stands for the pseudovector potential; and the cross stands for the electric charge current vertex.}
\label{fig2}
\end{figure}

Considering the lowest-order contribution shown in Fig. \ref{fig2}, we obtain the magnon-drag electric charge current $\bm j_\MAG^\E$ as
\begin{align*}
(\bm j_\MAG^\E)^\gamma
&\simeq
\frac{\I 2 D}{V}\sum_\QQ q^\gamma\omega_\QQ\int\frac{\D\Omega}{2\pi\hbar}\E^{-\I\Omega t/\hbar}\int\frac{\D\omega}{2\pi\hbar}
\notag\\
&
\times\Biggl(-\frac{\I e\hbar I^2}{2\pi mV}\Biggr)\sum_\KK k^\gamma\int\D E A_\mathrm{L}^\gamma(\Omega)
\notag\\
&\times
\Biggl[
g_{\KK\uparrow}(E_+)g_{\KK+\QQ\downarrow}(E+\omega)d_\QQ(\omega_+)d_\QQ(\omega_-)g_{\KK\uparrow}(E_-)
\notag\\
&
+
g_{\KK\downarrow}(E_+)g_{\KK-\QQ\uparrow}(E-\omega)d_\QQ(\omega_+)d_\QQ(\omega_-)g_{\KK\downarrow}(E_-)
\Biggr]^<.
\end{align*}
Through similar calculations as above, we obtain
\begin{align*}
\bm j_\MAG^\E
\simeq
-\frac{\D\bm A_\mathrm{L}(t)}{\D t}\sigma\frac{4I^2DP\delta\Delta}{3\alpha e(\Delta^2+\delta^2)^2V}\sum_\QQ\QQ^2\omega_\QQ\frac{\D n(\omega_\QQ)}{\D\omega_\QQ}.
\end{align*}
As a consequence, we can confirm that Onsager's reciprocity relation $S_\MAG=\Pi_\MAG/T$ is reproduced by using the relation $\D \bm A_\mathrm{L}(t)/\D t\to \bm\nabla T/T$ and the definition of the Seebeck coefficient $\bm j_\MAG^\E=:-\sigma S_\MAG\bm \nabla T$.

In summary,
we have shown that $\Pi_\MAG=C_\MAG T^{5/2}$ and $S_\MAG=\Pi_\MAG/T=C_\MAG T^{3/2}$ at a low temperature $T$ from the microscopic viewpoint, and that the coefficient $C_\MAG$ is proportional to the spin polarization $P$ of the electric conductivity.
Moreover,
from a comparison with experimental results for magnon-drag Peltier effects, we estimate that the strength of the electron--magnon interaction is about $0.3$ eV$\cdot\AA^{3/2}$ for permalloy.

\bibliographystyle{jpsj}

\end{document}